\documentclass[aps,twocolumn,superscriptaddress,showkeys,showpacs]{revtex4}

\usepackage{graphicx}
\usepackage{dcolumn}
\usepackage{bm}
\usepackage{color}
\usepackage{amsmath}

\begin{document}

\preprint{APS/123-QED}

\title{A Tale of Two Curricula:\\The performance of two thousand students\\
in introductory electromagnetism}

\author{Matthew A. \surname{Kohlmyer}}
	\altaffiliation[Current Address: ]{Department of Physics, North Carolina State University}
		\email{makohlmy@unity.ncsu.edu}
\affiliation{School of Physics, Georgia Institute of Technology, Atlanta, GA 30332}
\author{Marcos D. \surname{Caballero}}
\affiliation{Center for Nonlinear Science and School of Physics, Georgia Institute of Technology, Atlanta, GA 30332}
\author{Richard \surname{Catrambone}}
\affiliation{School of Psychology, Georgia Institute of Technology, Atlanta, GA 30332}
\author{Ruth W. \surname{Chabay}}
\affiliation{Department of Physics, North Carolina State University, Raleigh, NC 27695}
\author{Lin \surname{Ding}}
\affiliation{Department of Physics, The Ohio State University, Columbus, OH 43210 }
\author{Mark P. \surname{Haugan}}
\affiliation{Department of Physics, Purdue University, West Lafayette, IN 47907}
\author{M. Jackson \surname{Marr}}
\affiliation{School of Psychology, Georgia Institute of Technology, Atlanta, GA 30332}
\author{Bruce A. \surname{Sherwood}}
\affiliation{Department of Physics, North Carolina State University, Raleigh, NC 27695}
\author{Michael F. \surname{Schatz}}
\altaffiliation[Corresponding Author: ]{\texttt{michael.schatz@physics.gatech.edu}}
\affiliation{Center for Nonlinear Science and School of Physics, Georgia Institute of Technology, Atlanta, GA 30332}

\date{\today}

\begin{abstract}

The performance of over 2000 students in introductory calculus-based electromagnetism (E\&M) courses at four large research universities was measured using the Brief Electricity and Magnetism Assessment (BEMA).  Two different curricula were used at these universities: a traditional E\&M curriculum and the Matter \& Interactions (M\&I) curriculum.  At each university, post-instruction BEMA test averages were significantly higher for the M\&I curriculum than for the traditional curriculum.  
The differences in post-test averages cannot be explained by differences in variables such as pre-instruction BEMA scores, grade point average, or SAT scores.  BEMA performance on categories of items organized by subtopic was also compared at one of the universities; M\&I averages were significantly higher in each topic. The results suggest that the M\&I curriculum is more effective than the traditional curriculum at teaching E\&M concepts to students, possibly
because the learning progression in M\&I reorganizes and augments the traditional sequence of topics, for example, by
increasing early emphasis on the vector field concept and by emphasizing
the effects of fields on matter at the microscopic level.

\end{abstract}

\pacs{\textcolor{red}{01.40.Fk, 01.40.gb}}
\keywords{\textcolor{red}{Research in Physics Education, Curricula and evaluation, Teaching methods and strategies}}
\maketitle

\section{\label{sec:intro}Introduction}

Each year more than 100,000 students take calculus-based introductory physics 
at colleges and universities across the US. 
Such students must obtain a good working knowledge of introductory physics because physics concepts underpin the content of many advanced science and engineering courses required for the students' degree programs.  Unfortunately, many students do not acquire an adequate understanding of basic physics from the introductory courses; rates of failure and withdrawal in these courses are often high, and a large body of research has shown that student misconceptions about physics persist even after instruction has been completed \cite{hhcollege}. In recent years, there have been significant efforts to reform introductory physics instruction \cite{tutorialswash, eandmtipers, mazurpeer}.

Reforms of the course content (curricula) of introductory physics have not 
progressed as rapidly as reforms of content delivery methods (pedagogy). Most 
students are taught introductory physics in a large lecture format; the 
shortcomings of passive delivery of content in this venue are 
well-known \cite{nsf_shaping}.  A number of pedagogical modifications that 
improve student 
learning \cite{tutorialswash, eandmtipers, mazurpeer} have been 
introduced and are in widespread use; these modifications range from 
increasing active engagement of students in large lectures 
(e.g., Peer Instruction \cite{mazurpeer}and the  use 
of personal response system ``clickers'' \cite{wieman_clickers}) to reconfiguring the instructional 
environment \cite{beichner}.  By contrast, most 
students learn introductory physics following a canon of topics that has remained largely unchanged for decades regardless of the textbook edition or authors.   As a result the impact of changes in introductory physics curricula on 
improving student learning is not well understood.

At many universities and colleges, the introductory physics sequence consists of a one semester course with a focus on Newtonian mechanics followed by a one semester course in E\&M. There exist a number of standardized multiple-choice tests that can be used to assess objectively and efficiently student learning in large classes of introductory mechanics; some of these instruments have gained widespread acceptance and have been used to gauge the performance of thousands of mechanics students in educational institutions across the U.S.  \cite{hake6000}.  By contrast, fewer such standardized instruments exist for E\&M and no single E\&M assessment test is widely used.  As a result, relatively few measurements of student learning in large lecture introductory E\&M have been performed.

In this paper we report measurements of the performance of 2537 students in introductory E\&M courses at four large institutions of higher education: Carnegie Mellon University (CMU), Georgia Institute of Technology (GT), North Carolina State University (NCSU), and Purdue University (Purdue).   Two different curricula are evaluated:  a traditional curriculum, which for our purposes will be defined by a set of similarly organized textbooks in use during the study \footnote{Textbooks used in the traditional E\&M courses at the time of evaluation for each institution are Knight's \textit{Physics for Scientists and Engineers} (GT), Tipler's \textit{Physics for Scientists and Engineers} (Purdue), Giancoli's \textit{Physics for Scientists and Engineers} (NCSU), and Young and Freedman's \textit{University Physics} (CMU).}  and the Matter \& Interactions (M\&I) \cite{mandi2} curriculum.  M\&I differs from the traditional calculus-based curriculum in its emphasis on fundamental physical principles, microscopic models of matter, coherence in linking different domains of physics, and computer modeling \cite{atomsajp, mechajp, computajp}.  In particular, M\&I revises the learning progression of the second
semester introductory electromagnetism course by reorganizing and
augmenting the traditional sequence of topics, for example, by
increasing early emphasis on the vector field concept and by emphasizing
the effects of fields on matter at the microscopic level \cite{eandmajp}. 
Student performance is measured using the Brief Electricity and Magnetism Assessment (BEMA) a 30-item multiple choice test which covers basic topics that are common to both the traditional and M\&I electromagnetism curriculum including basic electrostatics, circuits, magnetic fields and forces, and induction \footnote{For a copy of the BEMA, contact the corresponding author.}. 
In the design of the BEMA, many instructors of introductory and advanced E\&M courses were asked to judge draft questions to ensure that questions included 
on the test did not favor one curriculum 
over another.  Moreover, careful evaluation of the BEMA suggests the test is reliable with adequate discriminatory power for both traditional and M\&I curricula \cite{dingBEMA}.

The paper is organized as follows:  In Section \ref{sec:summary} we present a 
summary of BEMA results across the four institutions which provides a snapshot of the performance measurements for students in both the traditional and M\&I curricula.  In Sections \ref{sec:gatech}-\ref{sec:cmu} we then discuss the detailed results from each individual institution in turn.  In Section \ref{sec:itemAnalysis} we analyze BEMA performance by individual item and topic, discussing possible reasons for performance differences, and we make concluding remarks and outline possible future research directions in Section \ref{sec:discussion}.

\section{\label{sec:summary}Summary of Common Cross-Institutional Trends}

\begin{figure}
	\centering
		\includegraphics[clip, trim=0mm 0mm 5mm 0mm]{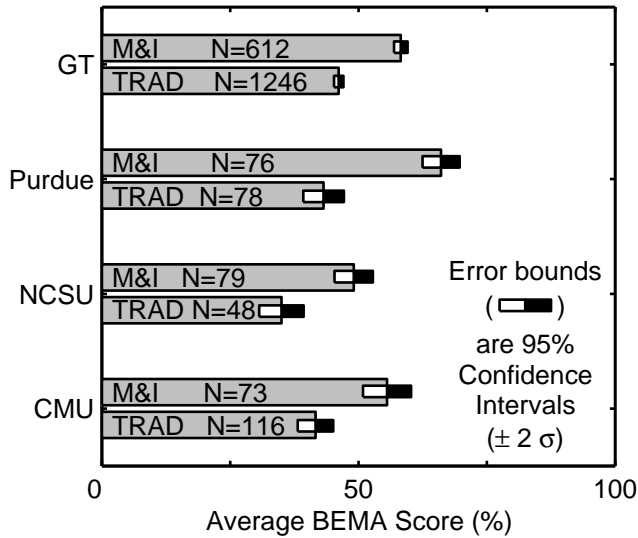}
	\caption{Average post-instruction BEMA scores at four academic institutions - The average BEMA test scores are shown for students who have completed a one-semester E\&M course with either the traditional (TRAD) or Matter \& Interactions (M\&I) 
curriculum.  
The number of students tested for each curriculum at each institution is indicated in the figure.  The error bounds represent the 95\% confidence intervals on the estimate of the average score.}
	\label{fig:summary_post}
\end{figure}

\begin{figure}
	\centering
		\includegraphics[clip,trim=2.5mm 0mm 5mm 0mm]{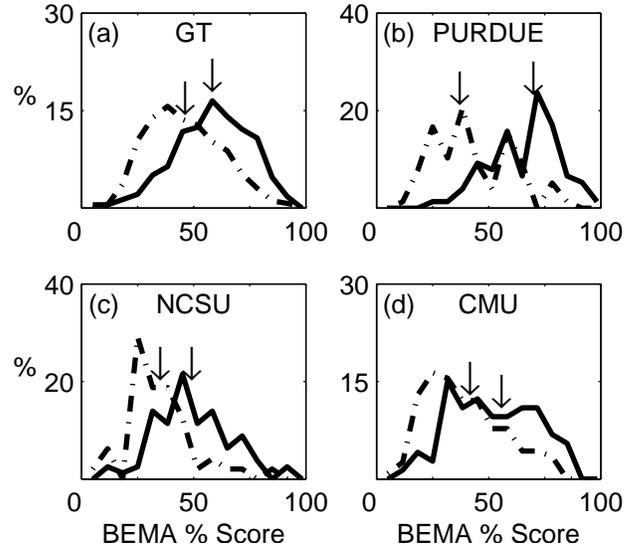}
	\caption{Post-instruction BEMA score distributions at four academic institutions - The percentage of students with a given BEMA test score is plotted for students who
 have completed an E\&M course with either a traditional (dot-dashed line) or
M\&I curriculum (solid line) at (a) GT, (b) Purdue, (c) NCSU, and (d) CMU.  The arrows indicate the location of the average score for each distribution. The right-most arrow in each subfigure corresponds to the M\&I course. The total number of students tested in each 
curriculum at each institution is the same as in Figure \ref{fig:summary_post}.  The plots are constructed from
binned data with bin widths equal to approximately 6.7\% of the maximum possible BEMA score (100\%).
}
	\label{fig:subplot_dist_line}
\end{figure}

Comparison of student scores on the BEMA at all four academic institutions 
suggests that students in the M\&I curriculum complete the E\&M course with a significantly better grasp of E\&M fundamentals than students who complete E\&M studies in a traditional curriculum (Figures \ref{fig:summary_post}, \ref{fig:subplot_dist_line}, \& \ref{fig:summary_norm_gain}). (A description of the methodology used to define ``significance'' is given in Appendix \ref{app:stats_method}.)   
Broadly speaking, the profiles of students at all institutions were similar; the vast majority of students in both curricula were engineering and/or natural science majors.  During the term, all students at a given institution were exposed to an instructional environment with similar boundary conditions on contact hours: large lecture sections that met for 2-4 hours per week (depending on the institution) in conjunction with smaller laboratory and/or recitation sections that typically met for 1-3 hours per week on average (again, depending on the institution).  We emphasize that, at a given institution, the contact hours were, for the most part, very similar for both M\&I and traditional courses (see Sections \ref{sec:gatech} - \ref{sec:cmu}).
Both the average BEMA scores (Figure \ref{fig:summary_post}) and the BEMA score distributions (Figure \ref{fig:subplot_dist_line}) were obtained at all institutions by administering the BEMA after students completed their respective E\&M courses.

A measure of the gain in student understanding as a result of instruction can be obtained by also administering the BEMA to students as they enter the course.   Specifically, the average increase in student understanding is measured by the average percentage gain, $G = O - I$, where $I$ is the average BEMA percentage score for students entering an E\&M course, and $O$ is the average end-of-course BEMA percentage score.  It has become customary \cite{hake6000} to report an average normalized gain $g$, where $g = G/(100 - I)$ and $(100 - I)$ represents the maximum possible percentage gain that could be obtained by a class of students with an average incoming BEMA score of $I$.   For $g$ reported in Figure \ref{fig:summary_norm_gain}, the Georgia Tech and Purdue data are shown only for students who took the BEMA both upon entering and upon leaving their E\&M course.
For the NCSU and CMU students in this study, $I$ was not measured.  In these cases, we estimate $g$ using measurements of $I$ for other similar student populations at each institution (See Section \ref{sec:ncsu} and \ref{sec:cmu} for details on, respectively, the NCSU and CMU estimates.)   With these qualifications, the data 
(Figure \ref{fig:summary_norm_gain}) 
show at all four academic institutions that students receiving instruction in the M\&I curriculum show significantly greater gains in understanding fundamental topics in E\&M than students who received instruction in a traditional curriculum.  

As we will discuss later, students who get A's in the course do better on the
BEMA than those who get B's, who in turn do better than those who get C's.
Comparison of average BEMA scores for a given final course grade in E\&M at CMU,
NCSU, and GT suggests that, roughly speaking, M\&I students perform one letter
grade higher than students in the traditional-content course. For example, on
average an M\&I student with a course grade of B does as well on the BEMA as the
traditional-content student with a course grade of A.


In addition to the common features described here, the E\&M instructional and assessment efforts contained a number of details unique to each academic institution.  We discuss these details below (Section \ref{sec:gatech}-\ref{sec:cmu}).

\begin{figure}
	\centering
		\includegraphics[clip, trim=0mm 0mm 5mm 0mm]{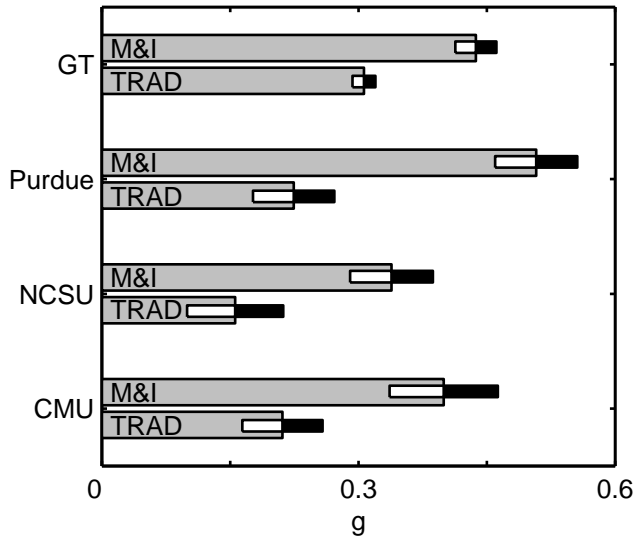}
	\caption{Gain in student understanding of E\&M at four academic institutions - The increase in student understanding resulting from a one-semester traditional (TRAD) or Matter \& Interactions (M\&I) course is 
measured using the average normalized gain $g$.   The number of students tested for each curriculum at each institution is:  GT M\&I: $N=297$, GT Trad.: $N=887$, Purdue M\&I: $N=76$, 
Purdue Trad.: $N=79$, NCSU M\&I: $N=79$, NCSU Trad.: $N=48$, CMU M\&I: $N=73$, CMU Trad.: $N=116$.  The error bounds represent
the 95\% confidence intervals on the estimate of the normalized gain.   The estimates of $g$ require the average BEMA scores for 
incoming students $I$; for the NCSU and CMU results, $I$ was computed differently than for the GT and Purdue results. (See Sections 
\ref{sec:summary},\ref{sec:ncsu} and \ref{sec:cmu} for details.)}
	\label{fig:summary_norm_gain}
\end{figure}

\section{\label{sec:gatech}Georgia Tech BEMA Results}

The typical introductory E\&M course at Georgia Tech is taught with 
three one-hour lectures per week in large lecture sections 
(150 to 250 students per section) and three hours per week in small group
(20 student) laboratories and/or recitations.  
In the traditional curriculum,  each student attends a two-hour 
laboratory and, in a separate room, a 
one-hour recitation each week; 
in the M\&I curriculum, each student meets once per week 
in a single room for a single three-hour session involving
both lab activities (for approximately 2 hours on average) 
and separate recitation activities (for approximately 1 hour on average).
The student population of the E\&M course (both traditional and M\&I) 
consists of 83\% engineering majors and 17\% science 
(including computer science) majors.

\begin{table}[h]
	\centering
	\caption{Georgia Tech BEMA test results are shown for five Matter \& Interactions sections (M$1$-M$5$) and eleven traditional sections (T$1$-T$11$).  Different lecturers are distinguished by a unique letter in column $L$. (Note that lecturer B in M$3$ was assisted by lecturer A.)  The average BEMA score $O$ for $N_O$ students completing the course is shown for all sections.  Moreover, in those sections where data is available,  the average BEMA score $I$ for $N_I$ students entering the course are indicated.
$N_m$ is the number of students in a given section who took the BEMA both at the beginning and at the end of their E\&M course. GPA is the incoming cumulative grade point average for students in a given section.}
		\begin{tabular}{|l||c|c|c||c|c|c|c|}
			\hline
			\textbf{ID}	& \textbf{L} & \textbf{N$_{\mathrm{I}}$}	& \textbf{I\%}	& \textbf{N$_{\mathrm{O}}$}	& \textbf{O\%} & \textbf{N$_\mathrm{m}$} & \textbf{GPA}\\
			\hline
			\textbf{M1}	& A	& 43	& 24.5 $\pm$ 2.3	& 40	& 59.8 $\pm$ 4.8 & 40 & 2.96$\pm$0.18\\
			\hline
			\textbf{M2}	& A	& n/a	& n/a	& 149	& 59.7 $\pm$ 2.8 & n/a& 2.99$\pm$0.10\\
			\hline
			\textbf{M3}	& B	& n/a	& n/a	& 146	& 57.4 $\pm$ 2.6 & n/a& n/a\\
			\hline
			\textbf{M4}	& C	& 138	& 27.7 $\pm$ 1.9	& 138	& 59.5 $\pm$ 2.7 & 132& 3.14$\pm$0.10\\
			\hline
			\textbf{M5}	& D	& 140	& 24.7 $\pm$ 1.4	& 139	& 55.9 $\pm$ 2.9 & 131& 3.07$\pm$0.09\\
			\hline
			\textbf{T1}	& E	& 231	& 22.9 $\pm$ 1.2	& 204	& 41.2 $\pm$ 1.9 & 180& 3.10$\pm$0.07\\
			\hline
			\textbf{T2}	& E	& 219	& 22.9 $\pm$ 1.3 	& 195	& 40.7 $\pm$ 1.9 & 176& 2.99$\pm$0.08\\
			\hline
			\textbf{T3}	& F	& 203	& 25.7 $\pm$ 1.4	& 136	& 51.9 $\pm$ 3.0 & 130& 3.01$\pm$0.09\\
			\hline
			\textbf{T4}	& F	& 212	& 25.1 $\pm$ 1.4	& 144	& 50.8 $\pm$ 2.5 & 133& 2.98$\pm$0.09\\
			\hline
			\textbf{T5}	& E	& n/a	& n/a	& 144	& 38.3 $\pm$ 2.5 & n/a& 3.09$\pm$0.08\\
			\hline
			\textbf{T6}	& G	& n/a	& n/a	& 29	& 45.2 $\pm$ 6.5 & n/a& 2.98$\pm$0.12\\
			\hline
			\textbf{T7}	& G	& n/a	& n/a	& 36	& 44.5 $\pm$ 4.9 & n/a& 2.81$\pm$0.12\\
			\hline
			\textbf{T8}	& H	& 87	& 28.1 $\pm$ 2.0	& 73	& 54.8 $\pm$ 4.7 & 59& 2.97$\pm$0.13\\
			\hline
			\textbf{T9}	& J	& 112	& 26.5 $\pm$ 2.1	& 84	& 51.6 $\pm$ 3.7 & 75& 2.94$\pm$0.11\\
			\hline
			\textbf{T10}	& F	& 128	& 25.3 $\pm$ 1.6	& 103	& 50.3 $\pm$ 3.0 & 88& 3.04$\pm$0.09\\
			\hline
			\textbf{T11}	& F	& 127	& 25.8 $\pm$ 1.8	& 98	& 49.5 $\pm$ 3.3 & 82& 3.03$\pm$0.10\\
			\hline
		\end{tabular}
	\label{tab:SummaryOfGT}
\end{table}

Table \ref{tab:SummaryOfGT} summarizes the Georgia Tech BEMA test results for 
individual sections.   
In all traditional and M\&I sections, $N_O$ students in each 
section took the BEMA during the last week of class at the completion of the course,
typically during the last lecture or lab session.
Moreover, in the majority of both traditional sections (T1-T4, T8-T11)
and M\&I sections (M1, M4 \& M5), $N_I$ students in each section took
the BEMA at the beginning of the course during the first week of class, typically during the first lecture or lab section. 
$N_I$ for a given section is approximately equal to the number of students 
enrolled in that section, while $N_O$ is usually smaller than $N_I$, sometimes
substantially so (e.g., T$3$ and T$4$), due to the logistics of administering the test.  Thus, in each section, 
only those $N_m$ students  who took the BEMA both on entering and on 
completion of the course are considered for the purposes of computing both 
the unnormalized gain $G$ and the normalized gain $g$.  The BEMA was 
administered using the same time 
limit (45 minutes) for both traditional and M\&I students.  
M\&I students were given no incentives for 
taking the BEMA; they were asked to take the exam seriously and told 
that the score on the BEMA would not 
affect their grade in the course.  Traditional students taking the 
BEMA were given bonus credit worth up to a 
maximum of 0.5 \% of their final course score, depending in part on 
their performance on the BEMA.  A performance incentive for only 
traditional students would not be expected to contribute to 
poorer performance of traditional students relative to M\&I students, and,
therefore,  cannot explain the Georgia Tech differences in performance 
summarized by Figs. \ref{fig:summary_post} and \ref{fig:subplot_dist_line}.

\begin{figure}[h]
	\centering
		\includegraphics[clip, trim=2.5mm 0mm 0mm 0mm]{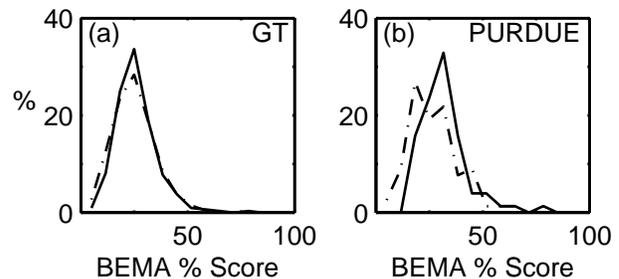}
	\caption{Pre-test BEMA score distributions for Georgia Tech and Purdue - The distributions of BEMA test scores for students before completing an E\&M course with either a traditional (dot-dashed line) or
M\&I curriculum (solid line) are shown for data from (a) GT ($\mathrm{N}$ = 1319 for traditional students, $\mathrm{N}$ = 321 for M\&I students) and (b) Purdue ($\mathrm{N}$ = 78 for traditional students, $\mathrm{N}$ = 76 for M\&I students). The plots are constructed from
binned data with bin widths equal to approximately 6.7\% of the maximum possible BEMA score (100\%).}
	\label{fig:subplot_dist_pre}
\end{figure}



\begin{figure*}[ht]
	\centering
		\includegraphics[clip, trim=0mm 0mm 5mm 0mm]{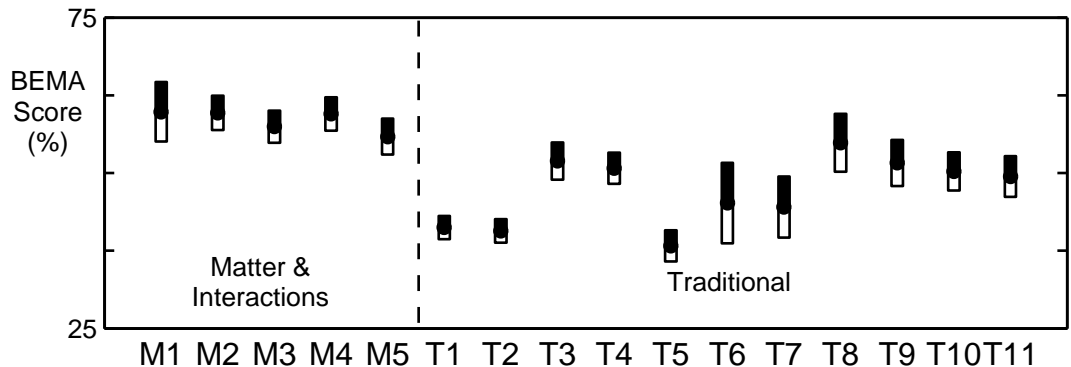}
	\caption{Average BEMA scores by section at Georgia Tech - The average end-of-semester BEMA scores for 11 traditional (T\#) and 5 M\&I (M\#) sections at GT are shown. The error bounds indicate 95\% confidence intervals on the estimates of the average for each section. 
The number of students tested in a particular section is given by $N_m$ in  Table \ref{tab:SummaryOfGT}.
}
	\label{fig:gt_by_section_ci}
\end{figure*}

\begin{figure}[h]
	\centering
		\includegraphics{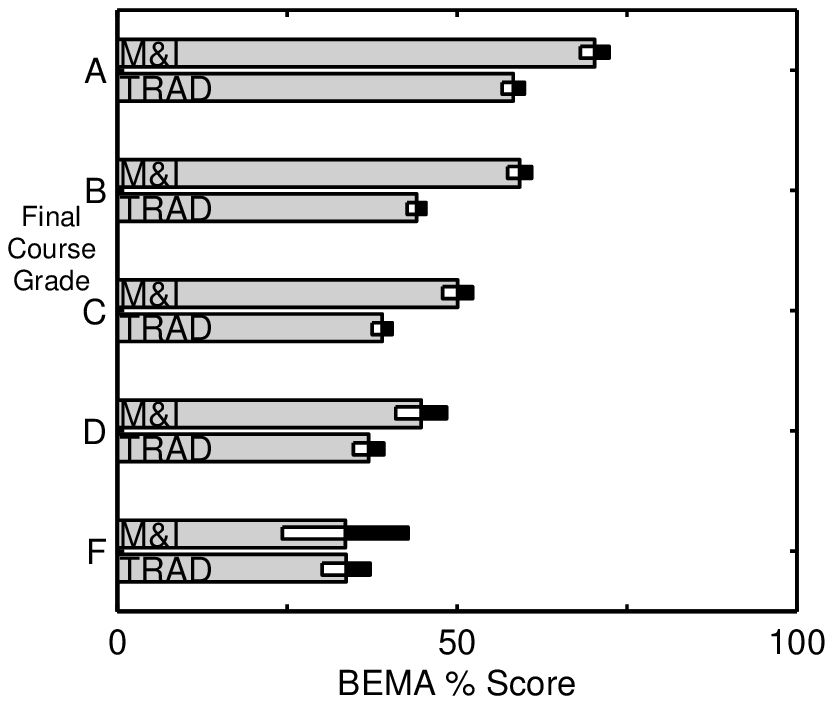}
	\caption{The average post-test BEMA score of all students receiving a particular final course grade in introductory E\&M at GT is shown. The error bounds indicate 95\% confidence intervals on the estimates of the average for each grade. The number of students for whom grades were obtained is $\mathrm{N}$ = 1233 for traditional students and $\mathrm{N}$ = 611 for M\&I students. 
}
	\label{fig:bema_vs_grade_gt}
\end{figure}

Figure \ref{fig:subplot_dist_pre}(a) demonstrates there was no significant difference between traditional and M\&I students 
in the distribution of 
pre-test scores on the BEMA.  The average pre-test score for all sections 
ranged from about 22\% to 28\%; a section-by-section comparison suggests 
there is no significant difference in pre-test scores on the BEMA between 
individual sections. (See Table \ref{tab:SummaryOfGT} and Appendix \ref{app:stats_method}).  
As an additional check on student populations
in the two curricula, we examined the students' grade point averages 
at the start of the E\&M courses; no significant difference in incoming GPA 
was found\footnote{The mean incoming GPA at Georgia Tech for M\&I students was 3.10 on a 4.0 scale, while the mean incoming GPA for traditional students was 3.11.}.  Thus, the student population
entering both courses is essentially the same.  Additionally, because the BEMA pre-test averages and 
the distribution of BEMA pre-test scores are essentially the same 
for the GT students in both curricula, we focus our remaining discussion 
on the post-test scores.

Figure \ref{fig:subplot_dist_line}(a) indicates the distribution of the BEMA 
post-test scores for the M\&I group is significantly different 
than the distribution for the traditional group.
Moreover, the BEMA post-test averages for each section 
(Figure \ref{fig:gt_by_section_ci}) suggest the 
M\&I sections consistently outperform the traditional sections.  
The M\&I BEMA averages across four different instructors are relatively 
consistent, while the BEMA averages of the traditional sections across five 
different instructors vary greatly. The use of Personal Response System (PRS)
``clicker'' questions may account for some of this difference. 
The lowest scoring sections 
(T1, T2 and T5 in Figure \ref{fig:gt_by_section_ci}) did not use 
clicker questions; by contrast, approximately 2-6 clicker questions 
were asked per lecture in all M\&I sections and all other traditional sections.
Nevertheless, even when the comparison between
sections is restricted to the traditional sections with the highest 
average BEMA scores (Sections T3, T4, T8 and T9, which were taught
by three different instructors who have a reputation of excellent teaching),
the M\&I sections demonstrated  significantly better performance 
(Appendix \ref{app:stats_method}).  

The data in Figure \ref{fig:bema_vs_grade_gt}
suggests a correlation between 
BEMA scores and final course grade at GT, with M\&I students outperforming
traditional students with the same final letter grade. Our finding that BEMA scores correlate strongly with final letter grade is not obvious. It seemed possible that the course grade was determined to a significant extent by the students' ability to work difficult multistep problems on exams, whereas the BEMA primarily measures basic concepts which, it was hoped, all students would have mastered. However, we find M\&I students exhibit a a one-letter-grade performance improvement as compared  with traditional students; specifically, the average BEMA scores are
statistically equivalent between traditional A students and M\&I B students, traditional B students and M\&I C 
students, and traditional C students and M\&I D students.  This difference in performance cannot be attributed to differences in the distribution of final grades; the percentage of students receiving a given final grade
in the M\&I sections (27.7\% As, 37.8\% Bs, 25.2\% Cs, 7.2\% Ds, and 2.1\% Fs) is similar to that in the traditional sections (29.8\% As, 34.4\% Bs, 24.3\% Cs, 8.8\% Ds, and 2.7\% Fs).






\section{\label{sec:purdue}Purdue BEMA results}
 
The curriculum comparison at Purdue focuses on an introductory E\&M course 
taught to electrical and computer engineering majors. The contact 
time was allocated somewhat differently for students in each curriculum; however, the total course contact time was similar for both traditional and 
M\&I students. Each week, traditional students met for three 50-minute large 
lectures (approximately 100 students per section) and two 50-minute 
small-group recitations (25-30 students); these students did not 
attend a laboratory.  M\&I students met for two 50-minute lectures per week in large lecture sections (approximately 100 students per section) and two hours per week in small group (25-30 students) laboratories. In addition, M\&I students attended a small group (25-30 students) recitation once a week for 50 minutes.  In all traditional and M\&I sections, students in each section took the BEMA during the last week of class at the completion of the course, typically during the last lecture or lab session. Moreover, students in each section took the BEMA at the beginning of the course during the first week of class, typically during the first lecture or lab section. 

Figure \ref{fig:subplot_dist_line}(b) indicates M\&I students 
significantly outperformed traditional students at Purdue. 
Students in both courses took the BEMA during a portion of a 
lab period with a 45-minute time limit for completion.
Both traditional and M\&I students took the assessment 
(both pre and post) in the same week. 
The ``initial state'' of the two groups upon entering their 
respective E\&M course was measured by 
comparison of the grade point averages between the two classes; no significant
difference was found\footnote{At Purdue, the mean incoming GPA for M\&I students was 3.25 and for traditional students the mean incoming GPA was 3.14.}.  Additionally, comparison of the distributions of the 
BEMA score upon entrance to the course shows only a small difference between
the two groups (Figure \ref{fig:subplot_dist_pre}(b)) that cannot 
account for the large post-test difference shown in 
Figure \ref{fig:subplot_dist_line}(b).




\section{\label{sec:ncsu}North Carolina State BEMA Results}

The introductory E\&M course at NC State is typically taught with 
three one-hour lectures per week in large lecture sections 
(about 80 students per section).  (Note, however, that one
M\&I section was taught in the SCALE-UP studio format \cite{beichner}.)  
In the traditional curriculum, each student attended a two-hour 
laboratory every two weeks; in the M\&I curriculum, each student attends 
a two-hour laboratory every week. 
Approximately three-fourths of the student population of the E\&M course (both traditional and M\&I) are engineering 
majors. 

One hundred twenty-seven volunteers were recruited from eight different 
sections (700 students total) by means of an in-class presentation made by a physics education research 
graduate student.  Students were paid \$15 for their participation in this out-of-class study. 
Prior to participation, students were told that they did not need to 
study for the test.  Just before the end of the semester, several testing 
times were scheduled to accommodate student schedules.  
The test was given in a classroom containing one computer per student, with a proctor present; 
each student took the test using an online homework system. 
Each student took the test independently with a 60-minute time limit. 



\begin{figure}[h]
	\centering
		\includegraphics[clip, trim=0mm 0mm 5mm 0mm]{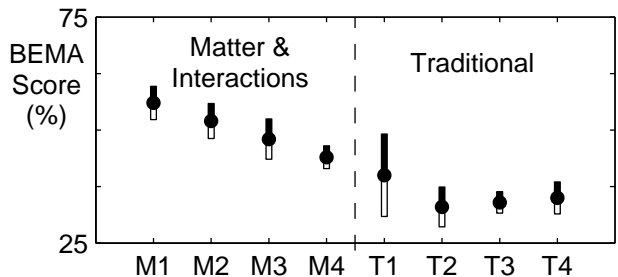}
	\caption{Average BEMA score by section at NCSU - The average end-of-semester BEMA scores for 4 traditional (T\#) and 4 M\&I (M\#) sections at NCSU are shown. 
The error bars indicate 95\% confidence intervals on the estimates of the average for each section. 
The numbers of students tested are:  $N=7$ for T1, $N=10$ for T2, $N=16$ for T3, $N=15$ for T4, 
$N=16$ for M1, $N=22$ for M2, $N=10$ for M3 and $N=31$ for M4. Note that section M4 was taught in the SCALE-UP studio format. 
}
	\label{fig:ncsu_by_section_ci}
\end{figure}

The difference in BEMA averages (shown in Figure \ref{fig:summary_post}) between the 
M\&I group and the traditional group is large and 
statistically significant as determined by the method outlined in Appendix \ref{app:stats_method}. Because students were recruited from eight different sections, it is of interest to observe how students from each section performed on the BEMA. 
Figure \ref{fig:ncsu_by_section_ci} shows the average scores of the individual sections for both M\&I and traditional 
groups.  Results of statistical tests (namely, the Kruskal-Wallis test\cite{nonparabook}) show that there was no significant difference in BEMA scores among the M\&I sections; similarly, no significant difference across the four traditional sections was detected. 
These results suggest that within each group students' BEMA scores were statistically uniform, and that the better performance of the M\&I students was not due to a few outlier sections that could have biased the results.

\begin{figure}[h]
	\centering
		\includegraphics[clip, trim=0mm 0mm 5mm 0mm]{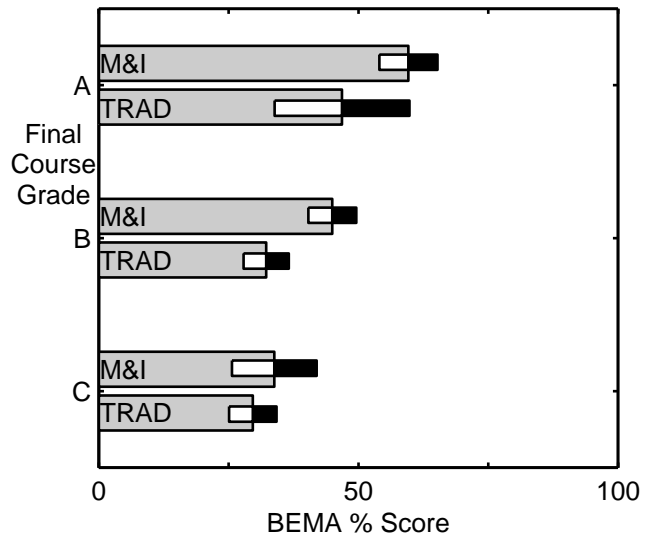}
	\caption{The average post-test BEMA score of all students receiving a particular letter grade in introductory E\&M at NCSU is shown. The error bounds indicate 95\% confidence intervals on the estimates of the average for each grade. The number of students for which grades were obtained is $\mathrm{N}$ = 48 for traditional students and $\mathrm{N}$ = 79 for M\&I students.}
	\label{fig:ncsu_gradesbema}
\end{figure}



One possible explanation of the results may be a recruitment bias; that 
is, higher-performing M\&I students and lower-achieving traditional 
students may have been recruited for the study.  To rule this out, 
participants' GPA, SAT scores as well as math and physics course grades
(prior to taking the E\&M course) were examined.  The two math courses 
from which students' grades were collected were the first and second 
semester of calculus courses; the physics course for which students' 
grades were collected was the calculus-based mechanics course. Using the method described in Appendix \ref{app:stats_method}, we found that there was no significant difference between the M\&I group and traditional group in any of these grades. Additionally, no significant difference was found in the SAT scores (verbal and math scores). These results suggest that the recruitment was not biased and that student participants from both the M\&I sections and traditional sections had similar academic backgrounds\footnote{At NCSU, the mean incoming GPA for M\&I students was 3.29 on a 4.0 scale, while the mean incoming GPA for traditional students was 3.23. The mean Calculus I GPAs were for 3.50 and 3.53 for M\&I and traditional students, respectively and the mean SAT scores were 1246.5 and 1248.3 for M\&I and traditional students, respectively.}



In the NCSU study, students were not given the BEMA prior to the start of their E\&M course.   However, a number of students from the same population, who 
were concurrently enrolled in introductory mechanics, did take the BEMA 
using via a web-based delivery system.  The average BEMA score of the mechanics students was 23\% \footnote{In fact, incoming students tend to earn similar BEMA scores across institutions regardless of curriculum: Georgia Tech (25.1\%), Carnegie Mellon (25.9\%) and Colorado (27\%).}. We use this value as an estimate for $I$ to compute the normalized gains shown in Figure \ref{fig:summary_norm_gain} which shows superior gain by M\&I students.



The data in Figure \ref{fig:ncsu_gradesbema} 
suggests a correlation between 
BEMA scores and final course grade at NCSU, with M\&I students outperforming
traditional students with the same final letter grade.  
Moreover, we find M\&I students exhibit a 
a one-letter-grade performance improvement; specifically,
the average BEMA scores are statistically equivalent between
traditional A students and M\&I B students.  Such a performance difference
might arise if fewer high final grades were awarded in M\&I than in the 
traditional course; under these circumstances, the A students in M\&I would
be more select and, perhaps, better than A students in 
the traditional course.  In fact, however, a somewhat larger percentage of 
higher final grades 
were earned in the M\&I sections 
(40.5\% As, 43.0\% Bs, 12.7\% Cs, 2.5\% Ds, \& 1.3\% Fs)
than in the traditional sections 
(25.0\% As, 54.2\% Bs, 18.7\% Cs, 2.1\% Ds, \& 0.0\% Fs).  Thus, the 
difference in performance on the BEMA cannot be attributed to differences 
in the distribution of final grades.

\section{\label{sec:cmu}Carnegie Mellon Retention Study}

The introductory E\&M course at Carnegie Mellon consisted of a large ($\sim$ 150 students) lecture that met three hours per week and a recitation section that met two hours per week; there was no laboratory component to this course.  For historical reasons, the course was separated into two versions: one for engineering majors that used the traditional curriculum and one for natural and computer science majors that used the M\&I curriculum.  The pedagogical aspects of 
both the traditional and M\&I courses were quite similar.

To probe the retention of E\&M concepts as a function of time, two
groups of students were recruited from each curriculum:  (1) Recent students 
of introductory E\&M, i.e., students who had taken the introductory E\&M  final exam 11 weeks prior to BEMA testing, and (2) ``old'' students of introductory E\&M, who had completed introductory E\&M  anywhere from 26 to 115 weeks prior 
to BEMA testing.  A total of 189 students volunteered for the study out of 
a pool of 1200 CMU students who had completed introductory E\&M at CMU and 
who were sent a recruitment email by a staff person outside the physics 
department.  
With a promise of a \$10 honorarium, the email asked for 
volunteers to take a retention test on an unspecified subject and stated that the test's purpose was to contribute to 
improvement in introductory courses.  The student volunteers 
took the BEMA during the evening in a separate proctored classroom. 
Just before taking the test, students were again told that they could help 
improve instruction at CMU by participating and doing their best; a poll of 
the students indicated, with one exception, that the volunteers arrived at the
examination room without knowledge of the test's subject matter. 
No pre-test was given to the students; however, an estimate of $I$, the average BEMA score prior to 
entering the E\&M course, was obtained by a separate study.  To obtain
this estimate, a different group of volunteers drawn from the 
appropriate pool of potential students for each curriculum, i.e., 
engineering students who had not yet taken the traditional E\&M course and
science students who had not yet taken the M\&I E\&M course.  These
volunteers were given the BEMA; we estimate $I$ = 28\% (N=14) 
for the traditional courses and $I$=23\% (N=10) for M\&I.


Disregarding the length of time since completing the E\&M course, it was found that the average BEMA score  $O$ = 41.6\% for students in the 
traditional curriculum is significantly lower than 
the $O$ = 55.6\% for students in the M\&I curriculum.
The participants from each course were not significantly different in 
background as measured by the average SAT verbal or math score.

\begin{figure}[h]
	\centering
		\includegraphics[clip, trim=0mm 0mm 5mm 0mm]{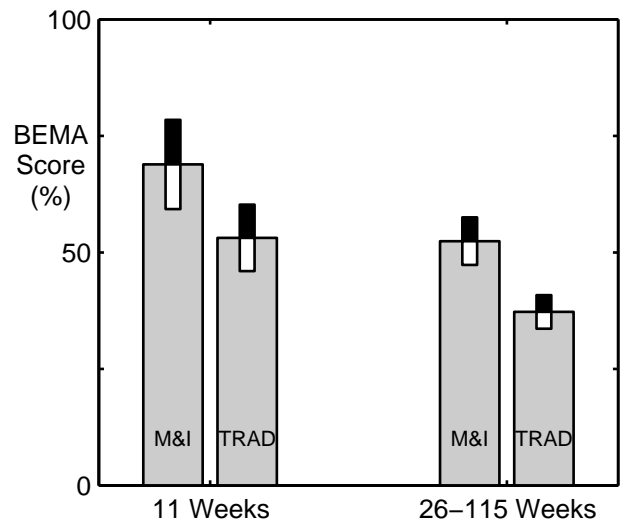}
	\caption{Retention of E\&M knowledge - Average scores on the BEMA vs time since completion of a course in introductory E\&M 
are shown for students at CMU from either a traditional curriculum (TRAD) or the Matter and Interactions (M\&I) curriculum.  The error bounds indicate 95\% confidence intervals on the estimates of the average for each section. 
The numbers of students tested were 116 for traditional and 73 for M\&I.
}
	\label{fig:time_cmu}
\end{figure}


Figure \ref{fig:time_cmu} shows that E\&M knowledge as measured by the BEMA showed a significant loss over the retention period for both M\&I and traditional students.  While the M\&I groups showed greater absolute retention at all grade levels than the traditional groups, the BEMA performances of students who most recently completed the E\&M course were also greater in the M\&I group.  The rate of loss in the two groups appeared to be the same, a result typically found in the experimental analysis of retention when comparing different initial ``degrees of learning'' \cite{slamecka,wixted}. Thus, as measured by BEMA performance we could not determine unequivocally that M\&I improved retention of E\&M knowledge over the traditional course beyond effects due to initial differences in performance on the BEMA. It's worth noting here that recent work has shown that 
better retention occurs for students exposed to improved pedagogical techniques\cite{pollock_longitudinal}. 

\begin{figure}[ht]
	\centering
		\includegraphics[clip, trim=0mm 0mm 5mm 0mm]{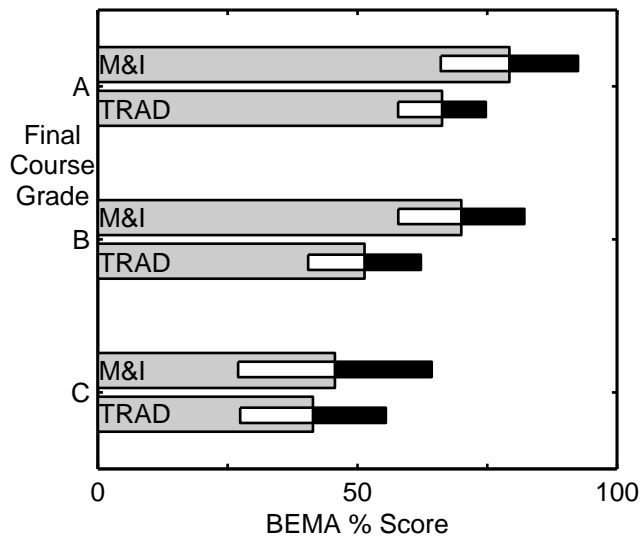}

	\caption{The average BEMA score for students receiving a particular final grade in introductory E\&M at CMU is shown. In all cases, the BEMA test
was administered 11 weeks after the completion of the course.  The error bounds indicate 95\% confidence intervals on the estimates of the average for each grade. The number of students for which grades were obtained is $\mathrm{N}$ = 14 for traditional students and $\mathrm{N}$ = 32 for M\&I students.}
	\label{fig:cmu_gradesbema}
\end{figure}

The data in Figure \ref{fig:cmu_gradesbema} suggests a correlation between BEMA scores and final course grade at CMU, with M\&I students outperforming traditional students with the same final letter grade.  Moreover, we find M\&I students exhibit a one-letter-grade performance improvement; specifically, the average BEMA scores are statistically equivalent between traditional A students and M\&I B students. Such a performance difference
might arise if fewer high final grades were awarded in M\&I than in the 
traditional course; under these circumstances, the A students in M\&I would
be more select and, perhaps, better than A students in 
the traditional course.  In fact, however, 
a somewhat larger percentage of higher final grades were earned in the M\&I sections (34.3\% As, 39.7\% Bs, 21.9\% Cs, 4.1\% Ds, and 0\% Fs) than in the traditional sections (25.0\% As, 37.9\% Bs, 31.9\% Cs, 5.2\% Ds, and 0.0\% Fs).
Thus, the 
difference in performance on the BEMA cannot be attributed to differences 
in the distribution of final grades.


\section{\label{sec:itemAnalysis}Item Analysis of the BEMA}

\begin{figure}
	\centering
		\includegraphics[clip, trim=0mm 0mm 5mm 0mm]{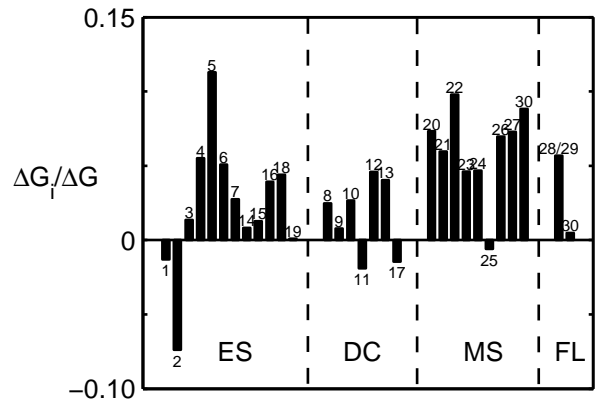}
	\caption{Fractional difference in performance for E\&M subtopics - The fractional difference in performance 
$\frac{\Delta G_i}{\Delta G}$ between M\&I and traditional students at GT is shown for each question on the BEMA. 
Positive (negative) $\frac{\Delta G_i}{\Delta G}$ indicates superior performance by M\&I (traditional) students.  
The numerical labels indicate the corresponding question number in order of appearance on the BEMA. The 
$\frac{\Delta G_i}{\Delta G}$ are grouped together into one of four topics:
Electrostatics (ES),  DC circuits (DC),  Magnetostatics (MS), or 
Faraday's Law and Induction (FL).}
	\label{fig:pca}
\end{figure}

We have seen superior performance on the BEMA from M\&I introductory E\&M classes 
as compared to traditional E\&M classes across multiple institutions.  
One question that arises is whether this result can be explained by M\&I students 
performing better in any one topic or set of topics in the E\&M curriculum.  
Because the content of the BEMA spans a broad range of topics, we can examine this 
question by dividing the individual BEMA items into different categories and 
comparing M\&I and traditional course performance in the individual 
categories.  There are some subjective decisions to be made when categorizing 
the items based on content and concepts, including the number of 
categories, the particular concepts they encompass, and which items belong to 
which categories.  Furthermore, certain items may involve more than one 
concept and could potentially fall into more than one category.  We 
decided, for simplicity, to group the BEMA items into just four categories 
covering different broad topics, namely, electrostatics, DC 
circuits, magnetostatics, and Faraday's Law of Induction.  Each item was 
placed into one and only one category; refer to Figure \ref{fig:pca} for the items that comprise each category \footnote{Item 29 of the BEMA is not scored separately--item 28 and 29 together count as one question, and must both be answered correctly to receive credit.}. Note that this 
is an \textit{a priori} categorization based on physics experts' judgment 
of the concepts covered by the items; it is not the result of internal 
correlations or factor analysis based on student data.  Using these categories, we compared M\&I and traditional 
performance in each category.  We chose to analyze the data from GT 
only, because we had the largest amount of data for traditional and M\&I courses across a range of different lecture 
sections from this institution.

We define the difference in performance between the two curricula
as $\Delta G = G_M-G_T$ where $G_M$ and $G_T$ are the (unnormalized) gains for
the M\&I and traditional curricula, respectively.  In the same way, we can 
determine $\Delta G_i$, the difference in performance of the $i^{th}$ 
BEMA question;  $\Delta G_i$ is equal to the percentage of M\&I students that 
answered the $i^{th}$ question correctly minus 
the percentage of traditional students that answered the same question correctly. 
Using these quantities, we define $\frac{\Delta G_i}{\Delta G}$, the fractional difference in performance 
for the $i^{th}$ question. $\frac{\Delta G_i}{\Delta G}$ can be thought of as the fractional contribution of 
the $i^{th}$ question to $\Delta G$ since $\displaystyle\sum_{i} \frac{\Delta G_i}{\Delta G} = 1$.   For equal 
weighting in the BEMA score (the scoring method that we used), a given question will make an ``average'' contribution 
to $\Delta G$ when the magnitude of 
$\frac{\Delta G_i}{\Delta G}$ is approximately equal to the inverse of the number of test questions (0.033 for
the 30-question BEMA).  Thus, when the magnitude of $\frac{\Delta G_i}{\Delta G}$ is significantly 
greater than 0.033, the corresponding question yields a greater than average contribution to $\Delta G$.
In addition, the sign of $\frac{\Delta G_i}{\Delta G}$ is noteworthy; a positive (negative) 
$\frac{\Delta G_i}{\Delta G}$ corresponds to
an item where on average the M\&I students scored higher (lower) than
traditional students.  (This presumes $\Delta G > 0$, which is the case for our data.)

The plot of $\frac{\Delta G_i}{\Delta G}$ for all questions on the BEMA provides a kind of 
``fingerprint'' for comparing in detail the performance of M\&I and traditional students (Figure \ref{fig:pca}). 
We see that the M\&I course has positive $\frac{\Delta G_i}{\Delta G}$ 
for almost all questions on the BEMA, and more 
than half of the questions (16) have values of  $\frac{\Delta G_i}{\Delta G}$ greater than 0.033.
\footnote{These questions are 4,5,6,12,13,16,17,18,20,21,22,23,24,26,27 and 28/29.}.  The grouping of 
the BEMA questions by category permits one to visualize which topics 
contribute most strongly to the difference
in performance.  For example, the difference in performance in 
magnetostatics is striking, where nearly every question
in this category has $\frac{\Delta G_i}{\Delta G} > 0.033$; in fact M\&I student performance on magnetostatics alone accounts for more than half (55\%) of 
the difference in performance $\Delta G$ relative to 
traditional students.  The positive $\frac{\Delta G_i}{\Delta G}$ for
DC circuits is worthy of note, even though these questions account for 
only 12\% of $\Delta G$.  Qualitatively speaking, 
the M\&I course seeks to connect the behavior of circuits to the behavior of
both transient and steady-state fields; this focus is decidedly 
non-traditional.  By contrast, the DC circuit questions on the BEMA
are quite traditional, so it is tempting to think that 
the traditional course might provide better training for responding 
to such questions.  However,  Figure \ref{fig:pca} demonstrates that
in fact M\&I students outperform traditional students on traditional
DC-circuit questions.   Performance in electrostatics also generally favors
the M\&I course (28\% contribution to $\Delta G$); however, we see the 
performance on question \#2 significantly favors the traditional 
course. The topic of question \#2 is 
the computation of electric forces using Coulomb's law.  It is possible that
the difference is due to greater time spent in the traditional class on
electric forces between point charges at the beginning of the course.  The
M\&I curriculum also discusses forces on charges, but moves into a full
discussion of electric fields due to point charges more quickly than the
traditional course, thereby devoting less time to discussing forces
exclusively.  By contrast, we also see the largest single percentage 
difference in favor of M\&I in question \#5, which deals with the direction 
of electric field vectors due to a permanent electric dipole.  The 
electric dipole plays an important role in the M\&I curriculum due to the 
curriculum's emphasis on the effects of electric fields on solid matter and polarization, topics which are often skipped or de-emphasized in the traditional course; this particular result is therefore not particularly surprising.  As a final note, the large values of $\frac{\Delta G_i}{\Delta G}$.  between M\&I 
and traditional courses in both magnetostatics and Faraday's Law are 
interesting because these topics are regarded as the most difficult for 
students due to their high level of abstraction and geometric complexity.  
It is therefore striking that the M\&I curriculum seems to be making the 
largest impact on the hardest topics, at least at Georgia Tech.

As an independent check on the significance of our item analysis, we used the method of contingency tables as described in Appendix \ref{app:stats_method} to compare the M\&I and traditional students' average scores in each individual category.  Here, a student's score in a category is computed as the sum of correct items in that category, where the number of items in the four categories range from 2 to 12.  The discrete nature of the data, as well as the non-normality and unequal variances of the distributions, make contingency tables the appropriate choice for this type of analysis. On the pre-test, we found no significant association between course treatment (M\&I versus traditional) and overall BEMA score on any category.  In contrast, the results of the contingency table analysis (see Appendix \ref{appsub:ctable}) for the post-test scores show significant association of BEMA score with treatment in each category. 
We interpret this as showing better performance across topics for students in the M\&I course.


\section{\label{sec:discussion}Discussion}

We have presented evidence that introductory calculus-based E\&M courses that use the Matter \& Interactions curriculum can lead to significantly higher student post-instruction averages on the Brief E\&M Assessment than courses using the traditional curriculum.  The strength of this evidence is bolstered by the number of different institutions where this effect is measured and by the large number of students involved in the measurements.  We interpret these results as showing that M\&I is more effective than the traditional curriculum at providing students with an understanding of the basic concepts and phenomena of electromagnetism.  This interpretation is based on accepting that the BEMA is a fair and accurate measurement of such an understanding.  We believe this is a reasonable proposition with which most E\&M instructors would agree, given that the BEMA's items cover a broad range of topics common to most introductory E\&M courses.  However, the BEMA was designed to measure just this minimal subset of common topics.  There may be other topics in which traditional students would outperform M\&I because they are not taught or de-emphasized in the M\&I course, and vice-versa.  


The BEMA is not the only instrument to assess student understanding of E\&M concepts.  The Conceptual Survey of Electricity and Magnetism (CSEM) was also designed for such a purpose \cite{csem_ajp}. With the exception of electric circuits, omitted by the CSEM, both instruments cover similar topics; in fact, several items are common to both tests.  However, the CSEM contains questions involving field lines, a topic which is not covered in the M\&I curriculum (a justification for this omission is discussed in \cite{eandmajp}).   Recent work has shown the CSEM and BEMA to be equivalent measures for changes of pedagogy\cite{pollock_csemandbema}. Nevertheless, it would be interesting to use the CSEM in comparative assessments of traditional and M\&I courses to see if it gives results similar to the BEMA; several of us are planning to do this in future semesters.

A major research question raised by these results is how and why the M\&I curriculum is leading to higher performance on the BEMA.  The post-instruction BEMA results measure only the total effect of the content and pedagogy of the entire course; there is no way to tease out from these measurements the effects of any individual elements of a course.  While it is true that interactive instruction methods (clickers) were used in almost every M\&I class measured, they were also used in many of the traditional classes.  Recall that M\&I sections at Georgia Tech still outperformed traditional sections with the two instructors noted for excellent pedagogical techniques.  Overall performance differences are not likely to be explained by differences in overall time-on-task; the weekly classroom 
contact time was equivalent for both M\&I and traditional students at
two of the four institutions (Georgia Tech and Purdue).  Time-on-task
for specific E\&M topics may partially explain performance differences
like those shown in Figure \ref{fig:pca}.   Comparing the percentage of 
total lecture hours devoted to each topic at GT, we find the M\&I course 
spends significantly more lecture time than the traditional course 
on Magnetostatics (24\% vs 12\%); this is
consistent with the superior performance of M\&I students on this topic.  However, we find superior performance of M\&I students on Electrostatics, for which both courses spend nearly equal lecture time (36\% vs 38\%). In addition, we also find superior M\&I performance of topics where the M\&I course spends significantly less lecture time than the traditional course, namely, DC circuits (15\% vs 25\%) and Faraday's Law/Induction (6\% vs 11\%).  We conclude that topical time-on-task alone is insufficient
to account for performance differences on the BEMA.
 
It is possible that the revised learning progression offered by the M\&I
E\&M curriculum is responsible for the higher performance on the BEMA by
M\&I students.   For example, more time is spent exclusively on charges
and fields early in the course, laying conceptual groundwork for the
mathematically more challenging topics of flux and Gauss's law which are
dealt with later than is traditional. Also, magnetic fields are
introduced earlier than is traditional, giving students more time to
master this difficult topic.  Finally, M\&I emphasizes the effects of
fields on matter at the microscopic level.  In some of the traditional
courses discussed in this paper, dipoles and polarization are not
discussed.  


\begin{acknowledgments}
We would like to thank Deborah Bennett, Lynn Bryan, Dan Able and Melissa Yale for their assistance in collecting the Purdue BEMA data. We also thank Keith Bujak for his assistance in editing this manuscript. This work was supported generously by National Science Foundation's Division of
Undergraduate Education (DUE0618504, DUE0618519, and DUE0618647). 
\end{acknowledgments}

\appendix
\section{\label{app:stats_method}Hypothesis Testing and Confidence Intervals}
In this paper we have emphasized the use of error bounds to indicate the size of comparison effects, but we also sometimes mentioned statistical evidence for being able to state that some comparison was or was not statistically significant. In this appendix, we present the details of how ``significance'' is determined.
\subsection{\label{appsub:diff}Is there a difference?}
In educational research we often wish to compare two or more methods of instruction to determine if (and how) they differ from each other.  In our case, we attempt to address the question of whether instruction in Matter and Interactions (M\&I) results in better performance on a standard test of Electromagnetism (E\&M) understanding (i.e., the Brief Electricity and Magnetism Assessment or the BEMA) than instruction in the traditional course.  We have gathered, under various arrangements, scores on this test for the M\&I and traditional classes.  Do they, in fact, differ?  And just what do we mean by differ? In what ways?  We need a set of procedures to allow us to answer these kinds of questions under conditions of incomplete information.  Our information is incomplete for a variety of reasons.  While we are basically interested in possible differential outcomes (e.g., in BEMA test scores) of our two instructional treatments (M\&I and traditional), there are many factors that might affect the outcome, that is, obscure any real differences due to the treatments alone.  The classes may differ in the abilities of the students, in the particular qualities of instructors, or in methods of course performance evaluation.  We may address some of these concerns by attempts to equate various conditions though proper sampling as well as more directly assessing potential differences.
 
For simplicity, let us assume that we are drawing a random sample of size $2n$ from the same population, that is, from all possible physics students who could properly participate in this study, a very large number, $N$.  We then randomly assign our two treatment conditions to the sample, yielding two samples of size $n$.  We then differentially expose these two samples to our two curricula, M\&I and traditional, and obtain a distribution of scores for each.  Ideally, we were not restricted to samples of size $2n$ (where $2n << N$) randomly drawn from a parent population, but could subject that entire population to our differential treatments by randomly assigning the two treatments among all members of the population. Essentially dividing the original population into two equal sub-populations of size $N/2$.  If our treatments had no effect, then the two sub-population distributions of scores would be identical and indistinguishable from the parent population.  If, however, we could show that the two distributions differed, then we could say that our treatments produced two different populations.  By different, we mean that one or more parameters (e.g., mean, median, variance, etc.) of the populations differed from each other.  But how big a difference is a difference?  That question can be addressed by classical hypothesis testing procedures (i.e., statistical inference).

Of course, we have to be satisfied by sampling from a population to obtain estimates of population parameters, one illustration of incomplete information.  Measures, that is, functions defined on samples are called statistics.  The arithmetic mean of a sample of size $n$, for example, is the sum of the sample values divided by $n$.  The sample mean will then be an estimate of the population mean; the sample variance an estimate of the population variance, etc.  Obviously, the larger the sample size, the better the estimate of a population parameter.  If we draw multiple independent random samples and compute a statistic, we will obtain distribution of the sample statistic.  A sample statistic is a random variable and its distribution is called a sampling distribution.  Sampling distributions are essential to the procedures of statistical inference; they describe sample-to-sample variability in measures on samples.  For example, if we are interested in determining whether two populations differ in their means; let us assume they are otherwise identical, we may draw a random sample of size $n$ from each and compute the mean of that sample.  Each value is an estimate of their respective population mean, but each is also but one value drawn from a distribution of sample means.  If the two populations were, the same, then the two sample means would be just two estimates of the same population mean because they would have come from the same sampling distribution.  The closer the two values are, the more likely this is the case; the greater their difference, the more likely it is they come from different sampling distributions and thus from different populations.  ``More (or less) likely'' is a phrase calling for quantification and probability theory provides that through measures called test statistics.  These have specific sampling distributions that allow probabilities of particular cases to be determined by consulting standard tables or through statistical packages.  Common examples include the $z$ statistic, Student's $t$, Chi-square, and the $F$-distribution.  All of these distributions are related to the normal distribution. The $z$-statistic is standard normal; the $t$, Chi-square, and $F$ are asymptotically normal. With some exceptions, their applications assume normality of the sampled parent distribution, though they differ in robustness with respect to that assumption.  In a typical implementation, a test statistic or, more commonly, a ``statistical test'' is chosen based on the stated hypotheses and by considering assumptions made about population characteristics, sampling procedures, and study design.

Statistical inference involves testing hypotheses about populations by computing appropriate test statistics on samples to obtain values from which probability estimates of obtaining those values can be determined.  These lead either to accepting or rejecting an hypothesis about some aspect of a population.  Many hypotheses involve inferences about measures of central tendency (e.g., the mean) or dispersion (e.g., the variance).  Formal hypothesis testing is stated in terms of a null hypothesis, $H_0$, and a mutually exclusive alternative, $H_1$.  The null hypotheses is assumed true and is rejected only by obtaining, through appropriate statistical testing, a probability value (``p-value'') less that some pre-assigned value. This probability value, called a the level of significance or $\alpha$, is usually 0.05.  Of course, one could be wrong in rejecting (a ``Type I Error'') or accepting (a ``Type II Error'') the null hypothesis regardless of the $p$-value obtained.  A result is either statistically significant or not; there is no ``more'', ``highly'', or ``less'' significant outcome.

For example, we can test the null hypothesis that the population mean scores for the M\&I and traditional curriculum treatments are equal (i.e., the scores all come from a common population, assuming all other population parameters are equal):

\begin{subequations}
\begin{eqnarray}
H_0: \mu_{MI} = \mu_{T}\:(\mathrm{or},\:\mu_{MI} - \mu_{T} = 0) \\
H_1: \mu_{MI} \neq \mu_{T}\:(\mathrm{or},\:\mu_{MI} - \mu_{T} \neq 0)
\end{eqnarray}
\end{subequations}

In this case we are considering two populations from which we sample independently and for each pair of samples we calculate the sample means and then take their difference.  We then have a sampling distribution of sample mean differences.  If our $H_0$ is true, we would expect the mean of that distribution to be zero.

If the probability $p$ corresponding to the computed test statistic is less than a selected threshold, typically $p < 0.05$, then the hypothesis of equal means is rejected. We deem the difference statistically significant and infer that the two populations are statistically different.  By contrast, if $p$ corresponding to the computed statistic is greater than a pre-assigned value, then the hypothesis of equal means cannot be ruled out.  The null and its associated alternative hypothesis can be more specific, for example, the above alternative hypothesis could be $H_1: \mu_{MI} > \mu_{T}\:(\mathrm{or},\:\mu_{MI} - \mu_{T} > 0)$.

\subsection{\label{appsub:normal}Is it normal?}
Which test statistic is most appropriate?  As already indicated, this depends on a number of factors including sample size and the characteristics of the parent populations from which the samples are drawn.  Recall we perform our tests based on the sampling distributions of the particular statistics of interest.  If we are interested in testing hypotheses about differences in means, then we will be concerned about the sampling distribution of those differences.  How do the parent distributions from which our samples are drawn affect the sampling distribution?  If the parent distributions are normal, then the sampling distribution of differences in means will also be normal. The difference of means is a simple linear transformation of the parent distributions. What if the parent distributions are not normal?  There are tests for this, given our sample distributions, as we indicate subsequently. But the Central Limit Theorem states that if random samples of size $n$ are drawn from a parent distribution with mean $\mu$ and finite standard deviation $\sigma$, then as $n$ increases, the sampling distribution approaches a normal distribution with mean $\mu$ and standard deviation $\sigma/\sqrt{n}$. Hence normality of the parent distribution is not required.  This applies equally to the case of differences in sample means.  Thus, given large enough sample sizes, we might be tempted to directly use the $z$-statistic to test our hypotheses about means. However, this test assumes we know the population variances and we virtually never do.  We might then resort to the $t$-distribution in which we estimate population variances from our samples, but the $t$-distribution assumes normality of the parent distributions.  While the $t$-distribution tests are relatively robust with respect to this assumption, not all parametric tests we wished to perform on our data are.  Moreover, the $t$-distribution tests are not appropriate for samples drawn from skewed distributions \cite{kirk_stats, zhou_biostats}. As we show below, through an appropriate test we found the BEMA scores in our studies were likely drawn from non-normal and skewed population distributions. 

Fortunately, there are powerful distribution-free methods, often called ``non-parametric'' statistics, that place far fewer constraints on parent distributions.  So, to be both consistent and conservative we subjected all our data to statistical tests using these methods.  However, because our sample sizes were typically large, we were often able to take advantage of the Central Limit Theorem and thus ultimately make use of the normal distribution.

\subsection{\label{appsub:nonnormal}Is it not normal?}
In Figure \ref{fig:subplot_dist_line} we display the distributions of scores on the BEMA for the M\&I and traditional groups at the various institutions where our studies were conducted.  Are we justified in assuming normality of the parent populations given these sample distributions?  Our null hypothesis would be that each of these sample distributions reflects a population normal distribution with unknown mean and variance against the alternative hypothesis that the population distributions are non-normal.  The general method is a goodness-of-fit test originally developed by Kolmogorov and extended by Lilliefors to of an unspecified normal distribution \cite{conover_nonpara, hollander_nonpara}. The basic approach is to assess the difference between a normal distribution ``constructed'' from the data and the actual data.  The data consist of a random sample $X_1$, $X_2$, $\dots$, $X_n$ of size $n$ drawn from some unknown distribution function, $F(x)$.  Recall, a distribution function is a cumulation (i.e., an integral) of a probability distribution or density function.  The normal distribution function is the familiar ogive, the integral of the normal density function - giving the probability: $P (x \leq a) = F (a), (0 < F (x) <1)$.
Under the null hypothesis, $F(x)$ is a normal distribution function and we can estimate its mean and standard deviation from our data.  The maximum likelihood estimate for the mean, $\mu$, is 

\begin{equation}
 \bar{X} = \frac{1}{n}\sum^{n}_{i=1}X_i
\end{equation}

The standard deviation, $\sigma$, is estimated by:

\begin{equation}
 s = \sqrt{\frac{1}{n-1}\sum^{n}_{i=1}\left(X_i-\bar{X}\right)}
\end{equation}

These values allow us to specify our hypothesized normal distribution.  We can now ``construct'' our empirical distribution $S(x)$ by computing z-scores from each of our sample values, defined by

\begin{equation}
 Z_i = \frac{X_i - \bar{X}}{s}, i=1,2,\dots,n.
\end{equation}

Now, we draw a graph of $S(x)$ using these $Z_i$ values and superimpose the normal distribution function $F(x)$ from our estimated parameters.  The Lilliefors test statistic is remarkably simple:
\begin{equation}
 T_L = \mathrm{max}\left|F(x)-S(x)\right|;
\end{equation}

\noindent that is, the maximum vertical distance between the two graphs.  A table with this test statistic's $p$-values can be found in standard texts on non-parametric statistics \cite{hollander_nonpara} or from appropriate statistical packages.  For large samples ($n > 31$), the $p < 0.01$ $T_L$ value, for example, is determined as $1.035/d_n$ where $d_n = (\sqrt{n}-0.01 + 0.83/\sqrt{n})$.  Obtaining a value that large or larger leads to rejection of the null hypothesis of normality at the 0.01 level of significance.

Using this test, all the distributions shown in Figure \ref{fig:subplot_dist_line} were determined to be significantly different from normal. At the 0.05 level, the following test distributions were found to be non-normal: GT M\&I pre-test, GT traditional pre-test, GT M\&I post-test, GT traditional post-test, Purdue M\&I pre-test, Purdue traditional pre-test, Purdue M\&I post-test, Purdue traditional post-test, NCSU traditional post-test, and CMU traditional post-test. Two test distributions, NCSU M\&I post-test and CMU M\&I post-test, were found to be normal. Demographic data was subjected to this test as well. The following demographic data were found to be non-normal at the 0.05 level: GT M\&I GPA, GT traditional GPA, GT M\&I E\&M grade, GT traditional E\&M grade, Purdue M\&I GPA, Purdue traditional GPA, NCSU M\&I GPA, NCSU traditional GPA, NCSU M\&I SAT score, NCSU traditional SAT score, NCSU M\&I Math and Physics GPA, NCSU traditional Math and Physics GPA, CMU M\&I SAT score, and CMU traditional SAT score. Given these results, we elected to adopt statistical tests that did not assume normality.

\subsection{\label{appsub:twosample}The two-sample tests and assumptions about variance}
As already discussed in the initial section of the Appendix, a number of our questions involved comparisons between M\&I and traditional treatments under various conditions.  In standard parametric statistics, hypotheses testing of such comparisons makes assumptions about variances in the populations under test.  For example, $t$-tests of differences in means used with two independent samples assume, in addition to normality, that the population variances are equal.  Likewise, in analysis of variance (ANOVA) tests of differences between means with $k$ independent samples ($k > 2$) also assume equal variances in the populations under test.  This assumption is called homogeneity of variance.  

Curiously, assumption of equal variances also extends to typical distribution-free methods testing hypothesis about differences between or among treatments \cite{conover_nonpara, hollander_nonpara}.  Thus, before applying such tests, we tested the hypothesis of equal variance.  In all cases tested we were unable to reject the null hypothesis of equal variances.  The variance tests we used are based on ranks and are akin to distribution-free methods for testing differences between group means (or medians).  Because the latter two-sample tests are easier to describe, we begin with hypothesis testing about differences between groups in measures of central tendency.  A brief description of the variance tests will follow.  Aside from the assumptions about equal variances, the tests to be described only assume random samples with independence within and between samples.

We have two samples $X_i$ ($i=1\dots m$) and $Y_i$ ($i=1\dots n$) for a total of $m + n = N$ observations.  For example, the $X_i$'s could be scores on the BEMA from traditional classes and the $Y_i$'s BEMA scores from the M\&I classes.  Putative differences in measures of central tendency, whether referring to means or medians, are sometimes called location shifts. Assuming the distributions, whatever their shape, are otherwise equal, then changes in the mean or median of one (e.g., produced by an experimental treatment) merely shifts it to the right or left by some amount $\Delta$.  If we are interested in differences in means, then 

\begin{equation}
\Delta = E(Y) - E(X), 
\end{equation}

\noindent the difference in expected values of the distributions is a measure of treatment effect.

Let $F(t)$ be the distribution function corresponding to population of traditional students and $G(t)$ the distribution function corresponding to population of M\&I students.  Our null hypothesis tested is

\begin{equation}
H_0: F(t) = G(t),\:\mathrm{for\:every\:t.} 
\end{equation}

That is, 

\begin{equation}
H_0: \Delta = 0 
\end{equation}

The alternatives are

\begin{equation}
H_1: F(t) \neq G(t) 
\end{equation}

or

\begin{equation}
H_1': G(t) = F(t + \Delta),\:\mathrm{for\:every\:t}.\:(i.e., \Delta > 0) 
\end{equation}

These two alternatives reflect whether we are simply interested in showing any difference, for example, whether entering scores on the BEMA for the M\&I and traditional groups differ; or, as in $H_1'$, whether post-instruction BEMA scores for the M\&I group exceed those of the traditional group.

The Wilcoxon (or Mann-Whitney) test statistic, $W$, is based upon rankings of the sample values.  The procedure is simple: Rank the combined sample scores $N = m + n$ from least to greatest, then pick ranked observations from one of the samples in the combined set, say M\&I ranks, $R(Y_j)$, and sum them,

\begin{equation}
 W=\sum_j^nR(Y_j).
\end{equation}

There are methods for handling tied ranks that we will not discuss in detail here.  Having chosen a level of significance, $\alpha$, and the particular alternative hypothesis, tables of this statistic can be found in any text on non-parametric statistics or from standard statistical software packages.

If, as in our case, the sample sizes are large ($n >20$), then $W$ approaches the normal distribution and one can use the standard normal tables.  The large-sample approximation to $W$ is found from the expected value and variance of the test statistic $W$,

\begin{equation}
 E(W) = \frac{n(m+n+1)}{2}
\end{equation}

\begin{equation}
 var(W) = \frac{mn(m+n+1)}{12}.
\end{equation}

The large-sample version of the Wilcoxon statistic, $W_z$, is then

\begin{equation}
 W_z = \frac{W-E(W)}{\sqrt{var(W)}}.
\end{equation}

Under the null hypothesis, $W_z$  approaches the standard normal distribution $N(0,1)$, so we reject $H_0$ if $W_z \geq z_{\alpha}$ where $\alpha$ is our level of significance.

The comparison data from each institution shown in Figure \ref{fig:subplot_dist_line} were each tested for significant differences between post-instruction BEMA scores in the traditional and M\&I treatments ($H_1: \Delta > 0$).  In all cases, the M\&I groups were shown to outperform the traditional groups at the 0.05 confidence level.

In addition, the pre-instruction BEMA scores from Georgia Tech and Purdue shown in Figure \ref{fig:subplot_dist_pre} were tested for differences. In this case, we found we could not reject the null hypothesis for Georgia Tech, but we detected a significant difference in the Purdue populations. For a discussion Purdue pre-instruction differences, refer to Section \ref{sec:purdue}. Finally, we tested demographic data, as listed at the end of Section \ref{appsub:nonnormal} of this Appendix. Matched sets were compared, e.g. GT M\&I GPAs and GT traditional GPAs. We found that we were unable to reject the null hypothesis, $H_0: \Delta =0$, in each matched set. Hence, we conclude that the student populations at each institution are similar insofar as GPAs, SAT scores and grades in Physics and Calculus courses are concerned.

\subsection{\label{appsub:variance}Homogeneity of variance tests}

Perhaps the simplest test of homogeneity of variance is the squared-ranks test \cite{conover_nonpara}.  It is quite similar to the Wilcoxon test described in Section \ref{appsub:twosample}. Because it concerns variances, squaring certain values plays a role.  Recall the definition of the variance of a distribution as the expected value of $(X-\mu)^2$.  If the mean of the distribution is unknown, as discussed before, we estimate it from our sample.  In the case of testing equality of variances with two independent samples, we have one random sample of $m$ values, $X_1$, $X_2$, $\dots$, $X_m$, and another of size $n$, $Y_1$, $Y_2$, $\dots$, $Y_n$.

We now determine the absolute deviation scores of each value from their respective sample means,

\begin{equation}
 U_i=\left|X_i-\bar{X}\right|,i=1,\dots,m
\end{equation}

and

\begin{equation}
 V_j=\left|Y_j-\bar{Y}\right|,j=1,\dots,n.
\end{equation}

As in the Wilcoxon test, we obtain the ranks of the combined deviation scores, a total of 
$N= m + n$.  If there are no ties, then the test statistic is simply based on the squares of the ranks from one of the samples, say, the $V_j$'s (if there are ties, the expression is more complicated, \cite{conover_nonpara}):

\begin{equation}
 D = \sum_j^n\left[R(V_j)\right]^2.
\end{equation}

The null hypothesis,$H_0$, is $Var(X) = Var(Y)$ and the alternative, $H_1$ is $Var(X) \neq Var(Y)$.  This test is not affected by differences in means, because variances of distributions are not affected by location shifts.

If, as in our case, the sample sizes are large, this test statistic approaches the standard normal distribution,

\begin{equation}
 D_z = \frac{D-n(N+1)(2N+1)/6}{\sqrt{mn(N+1)(2N+1)(8N+11)/180}}.
\end{equation}

Because we are only interested in any difference regardless of direction, we reject $H_0$ if 

\begin{equation}
\left|D_z\right| \geq z_{\frac{\alpha}{2}}, 
\end{equation}

\noindent where $\alpha$ is our chosen significance level.  Modifications of this test can be used to test differences in variances among $k > 2$ samples \cite{conover_nonpara, hollander_nonpara}.  As already indicated, in all cases applied to our data, we could not reject the null hypothesis of equal variances at the 0.05 level. Data are compared using matched sets, e.g. GT M\&I Pre-test scores and GT M\&I Post-test scores or GT M\&I GPAs and GT traditional GPAs, etc.. The inability to reject the null hypothesis at 0.05 level applies to all matched sets listed in Section \ref{appsub:nonnormal} of this Appendix.

\subsection{\label{appsub:cis}Gaining Confidence Intervals}
The specification of confidence intervals for selected parameters of population distributions is a common alternative to formal hypothesis testing. Indeed, many researchers much prefer this method when it can be applied for reasons we will not explore here \cite{kirk_stats}. But, simply put, confidence intervals can provide a ``quick picture'' of bounds on a population parameter based on sampling distribution estimates. They allow one to see if putative differences between group treatments are worth considering. This derives from our obtaining bounds on estimates of population parameters, such as means, medians, or variances. For simplicity, let us assume we are sampling from a normal distribution with known variance, $\sigma^2$, and attempting to determine bounds on the population mean, $\mu$, by selecting samples of size $n$ and computing the sample mean, $\bar{X}$. The sampling distribution of

\begin{equation}
 z = \frac{\left(\bar{X}-\mu\right)}{\left(\sigma/\sqrt{n}\right)}
\end{equation}

\noindent is the unit normal distribution $N(0, 1)$.  If we were to randomly draw one $z$ statistic from this distribution then the probability that the obtained $z$ will come from the open interval ($-z_{0.025}, z_{0.025}$) is

\begin{equation}
 P\left(-z_{0.025} < z < z_{0.025}\right) = 1 - 0.05 = 0.95.
\end{equation}

\noindent These values define confidence limits of the 95\% confidence interval for the population mean based on random samples drawn from that population.  The 0.95 probability specification is called the confidence coefficient.  The probability expressed in terms of the sample statistic is then

\begin{equation}
 P\left(\bar{X}-z_{0.025}\frac{\sigma}{\sqrt{n}} < \mu <  \bar{X}+z_{0.025}\frac{\sigma}{\sqrt{n}}\right).
\end{equation}

\noindent More generally, we can find a two-sided $100(1-\alpha)$\% confidence interval for the mean, $\mu$,

\begin{equation}
 \bar{X}-z_{\frac{\alpha}{2}}\frac{\sigma}{\sqrt{n}} < \mu <  \bar{X}+z_{\frac{\alpha}{2}}\frac{\sigma}{\sqrt{n}}.
\end{equation}

Generally speaking the population variance is unknown. It may be estimated from the sample in which case, given certain assumptions below, the $t$-statistic is the more appropriate and the two-sided $100(1-\alpha)$ confidence interval is then:

\begin{equation}
  \bar{X}-t_{\frac{\alpha}{2},\nu}\frac{s}{\sqrt{n}} < \mu <  \bar{X}+t_{\frac{\alpha}{2},\nu}\frac{s}{\sqrt{n}}.
\end{equation}

\noindent where $s$ is the standard deviation estimate from the sample (see Section \ref{appsub:nonnormal} in this appendix) and $\nu$ is the degrees of freedom given by $\nu=n-1$. The degrees of freedom is smaller than $n$ because we are using the sample to estimate the standard deviation. The appropriate values of $t$ are found in standard tables \cite{bevington}. As $\nu$ grows large, the $t$-distribution approaches normal, so for values greater than about 120, the $z$-table may be used.

We need to emphasize that it is not true that for any given sample the probability, $\alpha$, is that the mean, $\mu$, lies within that sample.  Once $\bar{X}$ is specified, it is no longer a random variable; either $\mu$ lies in that interval, or it does not.  Keep in mind that the analysis derives from considering all possible random samples of size $n$ drawn from the population to yield a distribution of confidence intervals.  Ninety-five percent of those intervals will include $\mu$ within the limits of $\pm t_{0.025}(\sigma/\sqrt{n}$), but 5\% will not.

Parametric confidence-interval determinations (e.g., using the $z$ statistic or the $t$-distribution) based on assumptions of normality (or at least symmetry with large sample sizes) may not be appropriate when confronted with non-normal, asymmetric distributions of the sort we encountered in our study. However, as stated earlier, the $t$-distribution is relatively robust with respect to normality provided the distribution is not significantly skewed. We obtained a measure of skewness for our sample distributions and determined that our distributions did not significantly depart from symmetric \cite{zhou_biostats}. We thus used the $t$-statistic for all the determinations of 95\% confidence intervals shown in Figures \ref{fig:summary_post}, \ref{fig:summary_norm_gain}, \ref{fig:gt_by_section_ci}, \ref{fig:bema_vs_grade_gt}, \ref{fig:ncsu_by_section_ci}, \ref{fig:ncsu_gradesbema}, \ref{fig:time_cmu}, and \ref{fig:cmu_gradesbema}.

\subsection{\label{appsub:ctable}Using Contingency Tables}

An analysis of a group of items within a given set divulges the contribution made by those items to the overall set. This is the approach taken in the first part of Section \ref{sec:itemAnalysis}. Alternatively, one can ask whether there is an association between two variables in this set. Do we find an association between one variable (treatment) and another variable (performance) on a given topic? Contingency table analysis can describe whether an association between treatment and performance exists and the confidence level of that association. When using contingency table analysis, one understands that the $p$-values obtained are conservative as compared to those obtained using parametric tests\cite{numericalrecipes}.

The approach is to form a table of events. An event can be any number of countable items. In our case, it will be total score on a given topic tested on the BEMA. This section will provide an example using data from the Magnetostatics item analysis for Georgia Tech given in Section \ref{sec:itemAnalysis}. By separating the responders into their given sections, traditional versus M\&I, and counting each responder's overall score on a given topic, one has proposed a valid contingency table. This table appears as the middle two columns, O$_\mathrm{MI}$ and O$_\mathrm{TRAD}$, in Table \ref{tab:obs_c_table}. A valid contingency table requires that no responder is counted twice. One could not use individual items as the events as a responder may have gotten several different questions correct. Using the total number of correct items ensures that a responder is counted only once.

\begin{table}[h]
	\centering
	\caption{Observed counts for number of correct answers in the Magnetostatics (MS) topic for the BEMA post-test are shown for all Matter \& Interactions and all traditional sections at Georgia Tech. The total number of correct items is denoted by $N_\mathrm{C}$ (maximum: 9). The number of students with $N_\mathrm{C}$ correct answers appears in the column O$_\mathrm{MI}$ for Matter \& Interactions and O$_\mathrm{TRAD}$ for traditional. The sum of these columns appears in O$_\mathrm{T}$.}
	\begin{tabular}{|l||c|c||c|}
	\hline
	N$_\mathrm{c}$ & O$_\mathrm{MI}$ & O$_\mathrm{TRAD}$ & O$_\mathrm{T}$\\\hline
	0 & 7 & 45 & 52 \\\hline
	1 & 21 & 155 & 176\\\hline
	2 & 33 & 224 & 257\\\hline
	3 & 47 & 227 & 274\\\hline
	4 & 59 & 195 & 254\\\hline
	5 & 90 & 142 & 232\\\hline
	6 & 118 & 113 & 231\\\hline
	7 & 102 & 72 & 174\\\hline
	8 & 87 & 56 & 143\\\hline
	9 & 48 & 17 & 65\\\hline
\end{tabular}
\label{tab:obs_c_table}
\end{table}

\begin{table}[h]
	\centering
	\caption{Expected counts for number of correct answers in the Magnetostatics (MS) topic for the BEMA post-test are shown for all Matter \& Interactions and all traditional sections at Georgia Tech. The total number of expected correct items is denoted by $N_\mathrm{C}$ (maximum: 9). The expected number of students with $N_\mathrm{C}$ correct answers appears in the column E$_\mathrm{MI}$ for Matter \& Interactions and E$_\mathrm{TRAD}$ for traditional. The sum of these columns appears in E$_\mathrm{T}$.}
	\begin{tabular}{|l||c|c||c|}
	\hline
	N$_\mathrm{c}$ & E$_\mathrm{MI}$ & E$_\mathrm{TRAD}$ & E$_\mathrm{T}$\\\hline
0 & 17.13 & 34.87 & 52\\\hline
1 & 57.97 & 118.03 & 176\\\hline
2 & 84.65 & 172.35 & 257\\\hline
3 & 90.25 & 183.75 & 274\\\hline
4 & 83.66 & 170.34 & 254\\\hline
5 & 76.42 & 155.58 & 232\\\hline
6 & 76.09 & 154.91 & 231\\\hline
7 & 57.31 & 116.69 & 174\\\hline
8 & 47.10 & 95.90 & 143\\\hline
9 & 21.41 & 43.59 & 65\\\hline
\end{tabular}
\label{tab:exp_c_table}
\end{table}

After counting the events, labeled $N_{ij}$, the column and row sums for table are computed. Summing down the column,

\begin{equation}
 N_{i.} = \sum_j N_{ij}
\end{equation}

\noindent is equivalent to counting the total number of responders in each treatment in Table \ref{tab:obs_c_table}. While summing across the rows,

\begin{equation}
 N_{.j} = \sum_i N_{ij}
\end{equation}

\noindent is equivalent to counting the total number of responders with a given score regardless of treatment. These numbers appear in column O$_\mathrm{T}$ in Table \ref{tab:obs_c_table}. One can determine the total number of responders by summing all rows and columns,

\begin{equation}
 N = \sum_{i,j} N_{ij} = \sum_i N_{i.} =\sum_j N_{.j}.
\end{equation}

\noindent This is equivalent to summing up the entries in column O$_\mathrm{T}$ in Table \ref{tab:obs_c_table}. 

We are able to compute an expected value for the number of events, $n_{ij}$, and compare that expectation value to the actual count. If treatment has no effect on the scores - that is, if we cannot distinguish any association between the treatment and score, we expect that the fraction of events in a given row is the same regardless of treatment. We can propose the null hypothesis,

\begin{equation}
 H_0:\:\frac{n_{ij}}{N_{.j}}=\frac{N_{i.}}{N}\:\mathrm{or}\:n_{ij}=\frac{N_{i.}N_{.j}}{N}
\end{equation}
with the alternative hypothesis,
\begin{equation}
 H_1:\:\frac{n_{ij}}{N_{.j}}\neq\frac{N_{i.}}{N}\:\mathrm{or}\:n_{ij}\neq\frac{N_{i.}N_{.j}}{N}.
\end{equation}

Table \ref{tab:exp_c_table} illustrates these expected values for the Magnetostatics topic. The columns E$_\mathrm{MI}$ and E$_\mathrm{TRAD}$ contain the expected number of students with a given score, N$_\mathrm{C}$. We can do a quick comparison of rows between Tables \ref{tab:obs_c_table} and \ref{tab:exp_c_table}. This provides an interesting contrast of higher (lower) expectations and actual counts.

A more rigorous approach is to perform a chi-square analysis with this expectation value, $n_{ij}$. We calculate the chi-square statistic as follows,

\begin{subequations}
 \begin{eqnarray}
  \chi^2 = \sum_{i,j}\frac{\left(N_{ij} - n_{ij}\right)^2}{n_{ij}},\\
  \nu = (I-1)(J-1)
 \end{eqnarray}

\end{subequations}

where $\nu$ is the number of degrees of freedom in the chi-square analysis. The degrees of freedom is determined by the number of rows, $I$, and the number of columns, $J$ in our contingency table (in our example $I$ = 10, $J$ = 2, so $\nu$ = 9). One can compare the reduced form of this statistic, $\chi^2/\nu$, at a given confidence level, $\alpha$, to computed values given in relevant texts or using any statistical package \cite{bevington}. Our example yields $\chi^2$ = 322.46, so that $\chi^2/\nu$ = 35.83. The critical value, for which we find our reduced statistic to be above, is $\chi^2_{crit}/\nu$ = 1.880. The $p$-value for our observed reduced chi-square statistic is much less than 0.0001. This shows significant association between treatment and score for the Magnetostatics topic on the BEMA post-test.

After performing this analysis, we found no association between treatment and score for the BEMA pre-test at Georgia Tech at the $\alpha=0.05$ level (all $p$-values were moderate, $p$ $>$ 0.20). However, the BEMA post-test scores showed a significant association between score and treatment at the $\alpha = 0.05$ level. The higher mean values achieved by the M\&I treatment dictate that the M\&I course is more effective for all topics; Electrostatics ($p$ $<$ 0.001), DC Circuits ($p$ $<$ 0.001), Magnetostatics ($p$ $<<$ 0.0001) and Faraday's Law ($p$ $<<$ 0.0001).

\bibliography{bema}
\bibliographystyle{apsper}

\end{document}